\documentclass[aps,shownopacs,superscriptaddress,nofootinbib,preprint,11pt]{revtex4}
 \usepackage[section,subsection,subsubsection]{extraplaceins}
 \usepackage{hyperref}
 \usepackage{amsmath}
 \usepackage{float}
 \usepackage{graphics}
 \usepackage{graphicx}
 \usepackage{epsfig}
 \usepackage{psfrag}
 \usepackage{color}
 \usepackage{slashed}
 \usepackage{axodraw}
 
 \usepackage{amsfonts}
 \usepackage{amssymb}

 \newcommand*{\be}{\begin{equation}}
 	\newcommand*{\ee}{\end{equation}}
 \newcommand*{\bea}{\begin{eqnarray}}
 	\newcommand*{\eea}{\end{eqnarray}}

 \newcommand{\comment}[1]{}

 \newcommand{\cref}[1]{Chapter~\ref{c.#1}}

 \def\beq{\begin{equation}}
 	\def\eeq{\end{equation}}
 \def\bea{\begin{eqnarray}}
 	\def\eea{\end{eqnarray}}
 \def\ba{\begin{array}}
 	\def\ea{\end{array}}
 \def\bi{\begin{itemize}}
 	\def\ei{\end{itemize}}
 \def\be{\begin{enumerate}}
 	\def\ee{\end{enumerate}}
 \def\bc{\begin{center}}
 	\def\ec{\end{center}}
 \def\bt{\begin{table}}
 	\def\et{\end{table}}
 \def\btb{\begin{tabular}}
 	\def\etb{\end{tabular}}

 	\def\lsim{\raise0.3ex\hbox{$\;<$\kern-0.75em\raise-1.1ex\hbox{$\sim\;$}}}
 	\def\gsim{\raise0.3ex\hbox{$\;>$\kern-0.75em\raise-1.1ex\hbox{$\sim\;$}}}

 	\def\beq {\begin{equation}}
 		\def\eeq {\end{equation}}
 	\def\bea {\begin{eqnarray}}
 		\def\eea {\end{eqnarray}}
 	
 	\preprint{DAMTP-2016-3
         TIFR/TH/15-44
         }
 	
\begin{document}

 		\title{Warped $R-$Parity Violation}
 		\author{B.C. Allanach}
 		\email{b.c.allanach@damtp.cam.ac.uk}
 		\affiliation{DAMTP, CMS, Wilberforce Road, University of
                  Cambridge, Cambridge, CB3 0WA, United 	Kingdom}
 		
 		\author{A. M. Iyer}
 		\email{abhishek@theory.tifr.res.in}
 		\affiliation{Department of Theoretical Physics, Tata Institute of Fundamental Research, Homi Bhabha Road, Colaba, Mumbai 400 005, India}
 		
 		\author{K. Sridhar}
 		\email{sridhar@theory.tifr.res.in}
 		\affiliation{Department of Theoretical Physics, Tata Institute of Fundamental Research, Homi Bhabha Road, Colaba, Mumbai 400 005, India}
 		
\begin{abstract}
 We consider a modified Randall-Sundrum (RS) framework between the
  Planck scale and  the GUT scale. In this scenario, RS works as a
  theory of flavour and not as a solution to the  hierarchy
  problem. The latter is
  resolved by supersymmetrising the bulk, so that  the minimal
  supersymmetric standard model being the 
  effective 4-dimensional theory. Matter fields are localised in the bulk in
  order to fit fermion-mass and mixing-data. If $R$-parity
  violating ($\slashed{R}_p$) terms 
  are allowed in the superpotential, their orders of magnitude throughout
  flavour space are then
  predicted, resulting in rich flavour textures. 
  If the $\slashed{R}_p$ contributions to neutrino masses are somewhat
  suppressed, then lepton-number violating models exist which explain the
  neutrino oscillation data while not being in contradiction with current
  experimental bounds.  
  Another promising model is one where baryon number is violated and Dirac
  neutrino masses result solely from fermion localisation. We sketch the likely
  discovery signatures of the baryon-number and the lepton-number violating
  cases. 
\end{abstract}
 		\maketitle
 		\section{Introduction}
 		\label{section1}
 		The discovery of the Higgs boson by the CMS
                \cite{Chatrchyan:2012xdj} and ATLAS
                \cite{Aad:2012tfa,Chatrchyan:2013lba} collaborations at the
                Large Hadron Collider has validated the status of the Standard
                Model (SM) as the correct theory of nature at the electroweak
                scale. The existence of a fundamental scalar in the
                SM raises questions regarding the stability of the Higgs mass
                in the face of radiative 
                corrections. Supersymmetry emerges as one of the most exciting
                prospects to address this problem due to its renormalizable
                nature and its consistency with electroweak precision
                data. The model, however, introduces a number of additional
                parameters which necessitates the study of its
                phenomenological implications using simplified models. 
 		
 		However, supersymmetry in its minimal form \textit{viz} the minimal supersymmetric standard model  (MSSM) does not
                give an explanation of the disparate couplings of the Higgs
                boson to different generations of fermions. This is also
                referred to as the fermion mass problem. 
 		In supersymmetry this problem can be addressed by considering
                strong wave function renormalization  of the matter
                fields \cite{Nelson:2000sn,Dudas:2010yh}. This  is due to
                renormalization group (RG) running from some fundamental
                renormalization 
                scale (where the Yukawa as well as the soft mass parameters are
                anarchic) to a low scale where they develop a hierarchical
                structure due to renormalization effects. The different
                running can be accounted by the different anomalous dimensions
                of each matter field coupled to some conformal sector.  
 		After canonically normalizing the kinetic terms, the
                superpotential terms are given by 
 		\begin{eqnarray}
 			\mathcal{W}&=&\epsilon^{q_i+h_u+u_j}\left(\hat Y^{ij}_U+\hat A^{ij}_UX\right)Q_iH_uU_j+\epsilon^{q_i+h_u+d_j}\left(\hat Y^{ij}_D+\hat A^{ij}_DX\right)Q_iH_dD_j\nonumber\\ &+&\epsilon^{l_i+h_d+e_j}\left(\hat Y^{ij}_e+\hat A^{ij}_eX\right)L_iH_dE_j
 			\label{superpotential}
 		\end{eqnarray}
 		while the K\"{a}hler terms are given by
 		\begin{equation}
 			\mathcal{K}=\sum_{F=Q,U,D,L,E}F^\dagger F+\hat C_{ij}\epsilon^{f_i+f_j}X^\dagger XF^\dagger F
 			\label{kahler}
 		\end{equation}
 		where $i$ and $j$ are generation indices and quantities with
                hatted quantities denoting $\mathcal{O}(1)$ parameters. $X$ is the SUSY
                breaking spurion parametrized as $X=\theta^2~F$. 
 		The expansion parameter $\epsilon\sim 0.02$ while the `charges'
                $q_i,h_{u,d}$  can be considered to be anomalous dimensions of
                the matter field coupling to a strong sector. Alternatively,
                they can be considered to be charges of the field under an
                extended gauge group $U(1)_{FN}$ \cite{Froggatt:1978nt}.  
 		The fermion mass matrix is then given as
 		\begin{equation}
 		m_f\sim\epsilon^{q_i+h_u+u_j}\frac{v}{\sqrt{2}}
 		\end{equation}
 		where $v\sim 246$ GeV is the vacuum expectation value (VEV) of
                the Higgs field. 
 		The charges $q_i$ are determined with the requirement of reproducing the correct pattern of fermion mass and mixing angles.
 		Soft supersymmetry breaking terms are generated when $F$ terms
                attains a VEV giving rise to the gravitino mass
                $m_{3/2}=\frac{\langle F \rangle}{M_{Pl}}$.  
	The mechanism which fixes the fermion masses and mixing angles will
        also determine the soft supersymmetry-breaking mass parameters, for
        example the squark mass-squared terms:
	\begin{equation}
	\tilde m^2_{ij}\sim {\mathcal O}(\epsilon^{q_i+q_j}m^2_{3/2}),
	\end{equation}
	and similarly for the other family-dependent soft supersymmetry-breaking terms. 
	This lends a certain level of predictivity to the
        orders of magnitude of soft breaking terms.  
	
 		The terms in Eqs.~\ref{superpotential},\ref{kahler}
                conserve $R-$parity \cite{Farrar:1978xj,Dimopoulos:1981zb},
                which   is defined as
 		\begin{equation}
 		R=(-1)^{3B+L+2s}
 		\end{equation}
 		where $s$ is spin of a particle and $B(L)$ is its
                corresponding baryon (lepton) number (alternatively, the same
                terms conserve matter parity
                \cite{Sakai:1981pk,Dimopoulos:1981dw,Weinberg:1981wj}). 
 		While $R$ parity conservation has many useful features,
                predicting the stability of dark matter and 
                a stable proton,
                there is no {\em a priori}\/ reason for it to be 
                a symmetry of
                the lagrangian\footnote{Scenarios in which R parity
                  originated as a discrete remnant of some extended gauge
                  symmetries were considered in
                  \cite{Krauss:1988zc,Font:1989ai,Mohapatra:1986pj,Martin:1992mq,Martin:1996kn,Allanach:2003eb}.}. 
 		Thus in general, the super-potential terms in
                Eq.(\ref{superpotential}) can also be extended to include
                terms which violate baryon and lepton-number, and are referred
                to as R-parity violating ($\slashed{R}_p$) terms\footnote{For
                  a detailed review on 
                 ($\slashed{R}_p$) supersymmtery see
                 Ref.~\cite{Barbier:2004ez}.}. 
 		The most general $\slashed{R}_p$ terms are given by
\begin{eqnarray}
  \mathcal{W}^{\Delta L=1}_{\slashed{R}_p}&=&\frac{\epsilon^{l_i+l_j+e_k}}{2}\lambda_{ijk}L_iL_jE_k+\frac{\epsilon^{l_i+q_j+d_k}}{2}\lambda'_{ijk}L_iQ_jD_k+\epsilon^{l_i+h_u}\mu'_iL_iH_u,\\\nonumber
  \mathcal{W}^{\Delta B=1}_{\slashed{R}_p}&=&\frac{\epsilon^{u_i+d_j+d_k}}{2}\lambda''_{ijk}U_iD_jD_k,
  \label{RPVlagrangian}
 		\end{eqnarray}
where we have omitted the gauge indices.

Whenever $R-$parity violation is introduced, one wonders where the apparent
relic density of dark matter might come from, given that it appears to be
stable on cosmological scales, and any MSSM fields will decay much too
quickly. One obvious answer is that massive hidden sector matter,  might
provide dark matter. Unfortunately, this would result in {\em no}\/ direct or
indirect signals for dark matter detection. Another
possibility~\cite{Takayama:2000uz,Moreau:2001sr} is 
that the lightest supersymmetric particle is the gravitino, which has Planck
suppressed couplings anyway. With additional smallish $\slashed{R}_p$
violating couplings, it is possible that its lifetime is much longer than the
age of the universe, resulting in a good dark matter candidate. We leave this
aspect of the model building to a future paper.

 		The anomalous dimensions of the matter fields can also be considered dual to the parameter which controls the localization of the field in an extra-dimensional scenario with strong warping \cite{Gherghetta:2000qt,Contino:2004vy,Gherghetta:2010cj}.
 		In this paper we consider the effects of introducing all such terms in a supersymmetric model on a gravitational background with strong warping also referred at as Randall-Sundrum (RS) model \cite{Randall:1999ee}.  
 		In Section~\ref{section2} we briefly introduce the model and set it up to understand the phenomenology. 
 		We review the technique used to determine the RS-model parameters which fit the fermion-mass and mixing-data at the high scale. The mathematical expressions used to determine the soft- and $\slashed{R}_p$-parameters are presented.
 		In section~\ref{section3} we discuss the implications of
                introducing $\slashed{R}_p$ couplings on various low-energy
                processes. We find that if baryon-number and lepton-number
                violating 
                terms are simultaneously allowed, consistency with constraints
                from proton decay require a slightly fine tuned choice of
                $~10^{-4}$ in some undetermined parameters usually expected to
                be of $\mathcal{O}(1)$. 
 		We then proceed to discuss simplified cases where either
                baryon- or lepton-number is conserved separately. Scenarios
                with lepton-number violation present solutions where the
                neutrinos can be Dirac-like even in the presence lepton number
                violating  terms. 
                In each case, we briefly comment on the LHC
                phenomenology, before presenting our conclusions. 
 		
 		\section{GUT-scale Randall-Sundrum model}
 		\label{section2}
 		We consider the following modified version of the original
                setup referred to as `GUT-scale RS'~\cite{Marti:2001iw,Choi:2003di,Choi:2003fk,Dudas:2010yh,Brummer:2011cp,Iyer:2013axa}.
 		Like the original RS model, it is a model of single
                extra-dimension compactified on a $S_1/Z_2$ orbifold. The line-element is given as 
 		\begin{equation}
 			ds^2=e^{-2\sigma(y)}\eta_{\mu\nu}dx^\mu dx^\nu+dy^2
 		\end{equation}
 		where $\sigma(y)=k|y|$ with $k$ denoting the reduced Planck
                scale and $R\sim 1/k$ being the size of the extra spatial
                dimension $y$.  
 		There are two opposite tension branes at each of the orbifold fixed points, $y=0$ and $y=\pi R$. Assuming the  scale of physics at the $y=0$ brane to be  $M_{Pl}$, the effective scale induced at the brane at $y=\pi R$ is given by
 		\begin{equation}
 			M_{IR}=e^{-\sigma(\pi)}M_{Pl}=\epsilon M_{Pl}\sim M_{GUT}
 		\end{equation}
 		Thus, in comparison to the original proposal in
                Ref.~\cite{Randall:1999ee}, the warp factor in this case is
                much 
                larger and hence {\em ab initio}\/ the model is no longer a
                solution to the                 hierarchy problem. 
 		Hence, supersymmetry is introduced into the bulk. With the
                GUT-scale Kaluza Klein (KK) modes decoupled from the theory, the
                spectrum of the effective    4D theory is that of MSSM. 
 		
 		We assume the two Higgs doublets to be localized on the
                infra-red (IR) brane (i.e.\ on the GUT brane) while the matter
                and gauge multiplets are in the bulk. 
 		The expressions for the fermion mass matrices are
 		\begin{eqnarray}
                  (m_{u})_{ij} &=&v_{u}\,\hat
                  Y_{u_{ij}}f(c_{Q_i})f(c_{u_j}) \nonumber \\
                  (m_{d})_{ij} &=&v_{d}\,\hat
                  Y_{d_{ij}}f(c_{Q_i})f(c_{d_j}) \nonumber \\
                  (m_{e})_{ij} &=&v_{d}\,\hat
                  Y_{e_{ij}}f(c_{L_i})f(c_{e_j}) 
 			\label{fermionmass}
 		\end{eqnarray}
 		where $i$ and $j$ are generation indices, $v_u=\frac{v}{2} \sin \beta$
                and $v_d=\frac{v}{2} \sin \beta$ are
                the VEVs of the up-type and down-type Higgs' of the MSSM,
                respectively and
                $c_{Z_i}$ are the  
                dimensionless bulk mass parameters of the matter multiplets $Z
                \in \{ Q, u, d, L, e\}$.  
 		The corresponding zero-mode wave-function $f$ is defined
                to be~\cite{Chang:1999nh,Gherghetta:2000qt} 
 		\begin{equation}
                  f(c) =  \sqrt{\frac{1-2 c }{e^{(1-2 c)\pi k R}-1}}e^{(0.5-c)kR\pi}.
 			\label{wavefunction}
 		\end{equation}
 		Using Eq.~\ref{fermionmass} and choosing 
                $c_{Z_i}\sim \mathcal{O}(1)$ and $\hat Y_{{U,D,E}}^{ij}\sim \mathcal{O}(1)$, one can explain the observed hierarchy in the fermion masses and mixings~\cite{Chang:1999nh,Gherghetta:2000qt,Grossman:1999ra,Iyer:2012db,Iyer:2013axa}. 
 		
A SUSY-breaking spurion $X=\theta^2 F$ is introduced on the GUT brane and
IR brane-localized contact interactions are introduced between the SM fields
and the SUSY breaking spurion $X$. The
soft SUSY breaking terms are then generated when the $F$-term attains a VEV
and are given by
\begin{eqnarray}
 m^2_{H_{u,d}} &=&  \hat h_{u,d} ~m_{3/2}^2\nonumber\\
  ( m_{{\tilde Z}}^2)_{ij} &=& m_{3/2}^2~\hat \beta_{\tilde Z_{ij}}~f(c_{Z_i})f(c_{Z_j})\nonumber\\
  A_{U}^{ij}&=&m_{3/2} \hat A_{U}^{ij}f(c_{Q_i})f(c_{u_j}), \nonumber \\
  A_{D}^{ij}&=&m_{3/2} \hat A_{D}^{ij}f(c_{Q_i})f(c_{d_j}), \nonumber \\
  A_{E}^{ij}&=&m_{3/2} \hat A_{E}^{ij}f(c_{L_i})f(c_{e_j}), \nonumber \\
  m_\alpha&=&\hat g_\alpha m_{3/2},
  \label{soft}
\end{eqnarray} 
where quantities denoted with a hat are fundamental dimensionless parameters,
which we assume are $\mathcal{O}(1)$. Here, $m^2_{H_{u,d}}$ are the up- and
down- Higgs mass squared soft SUSY-breaking parameters, $(m_{\tilde
  Z}^2)_{ij}$ the soft SUSY breaking mass squared matrix for sfermion $\tilde
Z$, $A_{(U,D,E)}^{ij}$ the matrix of trilinear soft SUSY-breaking interactions
for the up-quark, down-quark and charged leptons and $m_i$ the $i^{th}$
gaugino mass (where $\alpha \in \{ 3,2,1 \}$ denotes the MSSM gauge group
$SU(3)$, $SU(2)_L$, $U(1)_Y$), respectively.

$R$-parity violating interactions are introduced on this brane
and so they are considered to be generated at this scale\footnote{An
  equivalent description would correspond to the Higgs doublets and the
  $\slashed{R}_p$ violating terms localized on the ultra-violet (UV)
  brane.}. Like the soft 
parameters, the effective four-dimensional (4D) $\slashed{R}_p$ parameters can
also be expressed 
in terms of the bulk wavefunction of the fields. 
The effective 4D $\slashed{R}_p$ violating superpotential in a warped background is written 
 		\begin{eqnarray}
 		\mathcal{W}^{\Delta L=1}_{\slashed{R}_p}&=&\int dye^{-3ky}\delta(y-\pi R)\left(\lambda^{(5)}_{ijk}L_iL_jE_k+\frac{1}{2}\lambda^{'(5)}_{ijk}L_iQ_jD_k+\mu^{(5)}_iL_iH_u\right)\\\nonumber
 		\mathcal{W}^{\Delta B=1}_{\slashed{R}_p}&=&\int dye^{-3ky}\delta(y-\pi R)\lambda^{''(5)}_{ijk}U_iD_jD_k.
 		\label{RPVlagrangianRS}
 		\end{eqnarray}
 		$\lambda^{(5)}_{ijk},\lambda^{'(5)}_{ijk},\lambda^{''(5)}_{ijk}$ are 5D $\slashed{R}_p$ couplings which have mass dimension $-1$. Performing a KK decomposition of the fields and retaining only the zero modes\footnote{Higher KK modes in this model have mass $\sim M_{GUT}$ and are decoupled.}, the effective 4D $\slashed{R}_p$ couplings are written as

 		\begin{eqnarray}
 			\lambda_{ijk}&=&\hat\lambda_{ijk} f(c_{L_i})f(c_{L_j})f(c_{E_k}) \nonumber\\
 			\lambda'_{ijk}&=&\hat\lambda'_{ijk} f(c_{L_i})f(c_{Q_j})f(c_{D_k})\nonumber\\
 			\mu_i&=&\mu\hat\mu_i f(c_{L_i})e^{-kR\pi}
 			\label{leptonnoviolating1}
 		\end{eqnarray}
 		for the $\Delta L=1$ terms. $\mu$ is of order the
                electroweak scale and is chosen here to be 100 GeV. $\hat
                \lambda_{ijk},\hat{\lambda'_{ijk}}, 
 		\hat\mu_i$ are dimensionless $\mathcal{O}(1)$ couplings, where
                as $\hat \lambda\equiv k\lambda^{(5)}_{ijk}$, $\hat
                \lambda'\equiv k\lambda^{'(5)}_{ijk}$ and $\hat{\mu}\equiv k\mu^{(5)}$. 
 		The $\Delta B=1$ $\slashed{R}_p$ couplings are
 		\begin{equation}
 			\lambda''_{ijk}=\hat\lambda''_{ijk} f(c_{U_i})f(c_{D_j})f(c_{D_k})
 			\label{baryonnoviolating}
 		\end{equation}
 		with $\hat \lambda''_{ijk}=k\lambda''^{(5)}_{ijk}$.
 		The supersymmetric parameters in
                Eqs.~\ref{soft},\ref{leptonnoviolating1} and \ref{baryonnoviolating} are determined using the set of same set of $c_i$
 		parameters that fit the fermion masses and mixing at the GUT
                scale using Eqs.~\ref{fermionmass} and
                \ref{wavefunction}. This gives an order-of-magnitude level of
                predictability  for this framework, as these high-scale
                parameters can be subsequently evolved to generate a characteristic spectrum at the low scale.
 		The set of $\mathcal{O}(1)$ parameters (which includes the
                $c_i$ parameters as well as the $\mathcal{O}$(1) Yukawa
                parameters $\hat Y_{U,D,E}^{ij}$) is determined by
                performing a $\chi^2$ fit of their GUT-scale values
                to the data~\cite{Iyer:2013axa}.  
 		The $\chi^2$ function is defined as
 		\begin{equation}
                  \chi^2=\sum_i\frac{\left( \mathcal{O}^\textrm{theory}_i
(\{ c_j\} ,\{  \hat Y_{U}^{ij}\} ,\{  \hat Y_{D}^{ij}\} ,\{  \hat Y_{E}^{ij}\} )-\mathcal{O}^\textrm{expt}_i\right)^2}{\sigma_i^2},
 		\label{chisq}
 		\end{equation}
where $O_i^\textrm{theory}$ denotes the theoretical
prediction for observable $O_i$, $O_i^\textrm{expt}$ denotes the empirical
central 
value and the experimental uncertainty is written $\sigma_i$. 
$i \in \{ m_u, m_d,m_c,m_s,m_t,m_b, |V_{CKM}|_{ij}^\textrm{ind}\}$ constitute the hadronic observables whereas
$i \in \{ m_e, m_\mu,m_\tau, |V_{PMNS}|_{ij}^\textrm{ind}\}$ constitute the leptonic observables.
Both are fit independently (`$^\textrm{ind}$' indicates that the absolute
values of a selection of {\em
  independent}\/ entries - the off-diagonal entries - of the CKM and PMNS matrices
are fit, respectively).  
 We refer the interested reader to
Ref.~\cite{Iyer:2013axa} for further details. 
 	
\begin{figure}[here]
  \centering
  \includegraphics[angle=270,width=12cm]{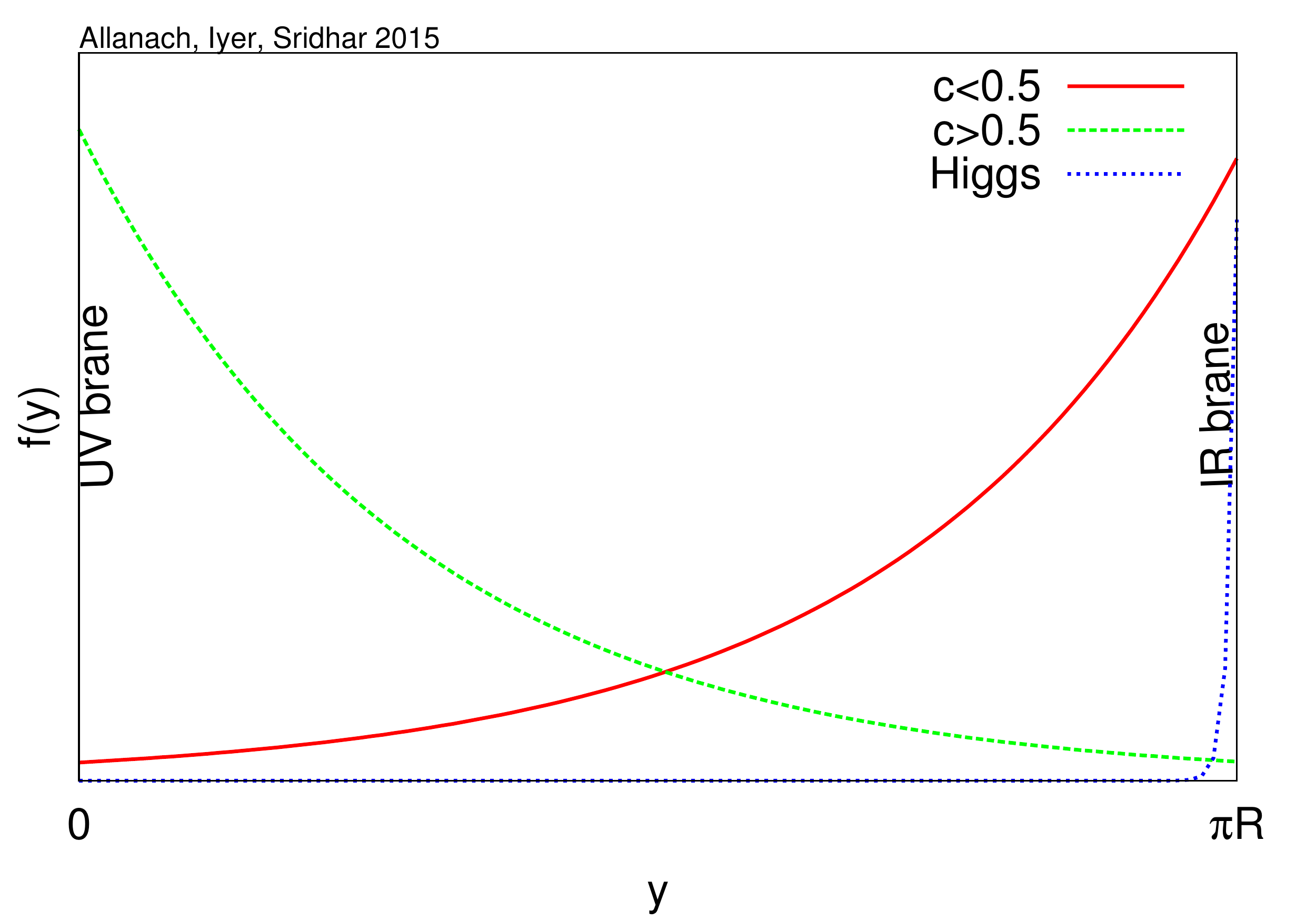}
  \caption{Localization of fermion profiles in the bulk depends upon the
    $c$ parameter. The Higgs is assumed to be strongly localized on the IR
    brane, as shown~\cite{sridhar}.} 
\label{localization}
\end{figure}	
Because of the small value of the warp factor $\epsilon\sim
0.02$, the $c_i$ parameters for the lighter generations (eg.\
the electron or the up and down quark) are close to $2.5$
while for the third generation of superfields containing a top, they are close
to 
-1. $c_i<0.5$ reflects a 
localization more on the IR brane (where the Higgs doublets are localized),
while $c_i>0.5$ localizes the superfield closer to
the UV brane, as heuristically depicted in Fig.~\ref{localization}. The $c_i$
parameters for all 
charged matter superfields except for $Q^3$ and $t_R$  are scanned in the range
$[0,\ 3.5]$, while $c_{Q^3}$ is scanned is scanned in $[0,\ 1.5]$ and
$c_{t_R}$ is 
restricted to $[-2.5,\ 0.5]$. This different choice for $c_{Q_3},c_{t_R}$ is to
facilitate a good fit to the top quark mass.
We remind the reader that fits to fermion mass and mixing data are done {\em
  independently}\ 
for the quark sector and the leptonic sector, since to a good
approximation (i.e.\ at tree-level), the two sectors are decoupled. 

% Figure \ref{CDF} gives the Cumulative distribution function for the $\chi^2$ function. This leads to a good fit $\chi^2<5$ for most of the points in both the hadronic and the leptonic sector.
 		
 		% \begin{figure}
                %   \begin{tabular}{cc}
                %     \includegraphics[width=8cm]{cdftb5.png}&\includegraphics[width=8cm]{cdftb10.png}
                %   \end{tabular}
                %   \caption{ }
                %   \label{CDF}
 		% \end{figure}
 		
The fits in the leptonic sector includes fitting the neutrino data by
means of introduction of three parameters $c_{N_i}$ corresponding to three
right handed neutrinos. To account for small neutrino masses at the sub eV
level,
$c_{N_i}$ 
are scanned in the range $[5.5,\ 7]$ in order to imbue Dirac neutrino masses via
Eq.~\ref{fermionmass}.
The presence of lepton-number violating operators gives rise to
additional Majorana contributions to the neutrino masses. By focussing on
regions of the 
parameter space where these contributions are suppressed, we will find
that the dominant contribution to the neutrino mass is from
Eq.~\ref{fermionmass} and hence are primarily of Dirac type. 
  		
The fits are performed for two separate values of $\tan\beta=5,10$. 
Smaller $\tan\beta$ facilitates a  localization of the light down sector
fields closer to the UV brane owing to a larger value of
$\cos\beta$ in 
the mass matrix in Eq.~\ref{fermionmass}. As we  shall explain later, this helps
in generating a smaller value for the $\slashed{R}_p$ couplings, enabling them
to satisfy experimental constraints more easily , which are typically upper
bounds.   		
The $\mathcal{O}(1)$ model parameters are determined by minimising the $\chi^2$ function in  Eq.~\ref{chisq}. 
The minimisation is performed by {\tt{MINUIT}} \cite{James:1975dr} which looks
for a minimum around a guess value of $c$ parameters and $\mathcal{O}$(1)
Yukawa parameters.  
The guess values are randomly generated in the ranges given above. This is
repeated for $10^5$ choices of guess values each constituting a separate
minimization. {\tt MINUIT}~has trouble searching our parameter space, and
finds many distinct local minima, depending upon which random guess we start
with. We view this as a sampling of the `good-fit' parameter space, and
all points which satisfy $\chi^2<10$ are accepted as being a reasonable
`fit'. We remind the reader that this is not a fit to data in the usual
statistical 
sense: rather it is a fit to the orders of magnitude of the masses and mixings. 
In addition to the $c_i$ parameters, the $R-$parity conserving hatted
$\mathcal{O}$(1) 
Yukawa parameters are all allowed to vary between 0.1 and 10, whereas the
hatted $\slashed{R}_p$ violating parameters are set to 1 and are not varied. 
With the $c_i$'s fixed in this manner, for each sampling, we predict the 
orders of magnitude 	of the $\slashed{R}_p$ parameters. 
 		
\section{$\slashed{R}_p$ parameters}
\label{section3}
We now focus on the distribution of the various $\slashed{R}_p$ couplings
which are determined from the fermion mass fits. As given in
Eq.\ref{RPVlagrangian}, $\slashed{R}_p$ terms include both
baryon-number and lepton-number violating interactions. The lepton-number
violating 
interactions include the trilinear couplings ($\lambda_{ijk},\lambda'_{ijk}$)
and the bilinear operators $\mu_i$.
$\lambda_{ijk}$ is anti-symmetric in $i \leftrightarrow j$ because of the
$SU(2)_L$ 
structure, as is $\lambda''_{kij}$ because of the implicit $SU(3)$ structure.
 		
On account of the introduction of the $\slashed{R}_p$ operators on the same brane as the Higgs superfields, their magnitude can be roughly understood from the generation indices in these couplings. For instance, consider
$\lambda_{111}$ which is a product of the zero mode profiles of some first-generation fermions and $\lambda_{333}$ is the corresponding product of third-generation fermions. 
Since, as Fig.~\ref{localization} illustrates, the lighter fermion generations
have a tendency 
to be 
localized away from the Higgs ($c_i>0.5$), the corresponding value of the
profile on the IR brane is small. The
third 
generation is relatively heavy and has a value of $c_i$ smaller than those
for the lighter generations. As a result the corresponding value of the
profile on the IR brane is relatively larger. This results in a larger value
for $\lambda_{333}$ as compared to $\lambda_{111}$.  
Similarly, $\mu_3\geq \mu_2\geq\mu_1$. In evaluating the $\slashed{R}_p$ 
parameters, the $\mathcal{O}(1)$ parameters
$\hat\lambda_{ijk},\hat\lambda'_{ijk}, \hat\lambda''_{ijk}$ and $\hat\mu_i$
were all chosen to be 1 (unless they are set to zero by requiring baryon or
lepton number conservation). We therefore should bear in mind that we provide
order of magnitude predictions, which will be multiplied by some order one
parameter. 

The predictions thus obtained are then filtered against
2 $\sigma$ upper bounds on $\slashed{R}_p$ violating parameters,
for instance from the non-observation of
$\mu\rightarrow e\gamma$ \cite{deGouvea:2000cf}, leptonic
decay of long lived neutral Kaon \cite{Choudhury:1996ia},
bounds from $n-\bar n$ oscillations and double neutron decay
\cite{Goity:1994dq} or constraints from the electroweak
precision 
tests \cite{Bhattacharyya:1994yc,Bhattacharyya:1995pr,ledroit}
\textit{etc}. A complete list of constraints on the various
$\slashed{R}_p$ parameters that we use is given in Tables 6.1 to 6.5 of
Ref.~\cite{Barbier:2004ez}, although in the first instance we do not apply
bounds from nucleon decay, upon which more later. The constraints do depend
upon the supersymmetric 
spectrum, for example  the branching ratio of $B \rightarrow \tau \nu$
\begin{equation}
\lambda'_{333} < 0.32 \left( \frac{m_{\tilde b_R}}{\textrm{100 GeV}}\right)
\end{equation}
depends upon the right-handed sbottom mass $m_{\tilde b_R}$.
We shall provide predictions for viable ranges of $\slashed{R}_p$ violating
parameters for soft masses $\tilde m \gtrsim 300$ GeV. 
If any one of the 2$\sigma$ bounds is violated in the case with all hatted
$\slashed{R}_p$ violating parameters fixed to one and 300 GeV sparticles, the
point is discarded.  After this filtering, we obtain 2203 good-fit
parameter points to the quark mass and mixing data and 848 to the
lepton mass and mixing data. We combine each set of good-fit quark parameters
with each set of good-fit lepton parameters (since they are approximately
independent, as explained above) in order to determine the possible ranges of
the various $\slashed{R}_p$ violating couplings. 

\begin{figure}
\includegraphics[angle=270,width=\textwidth]{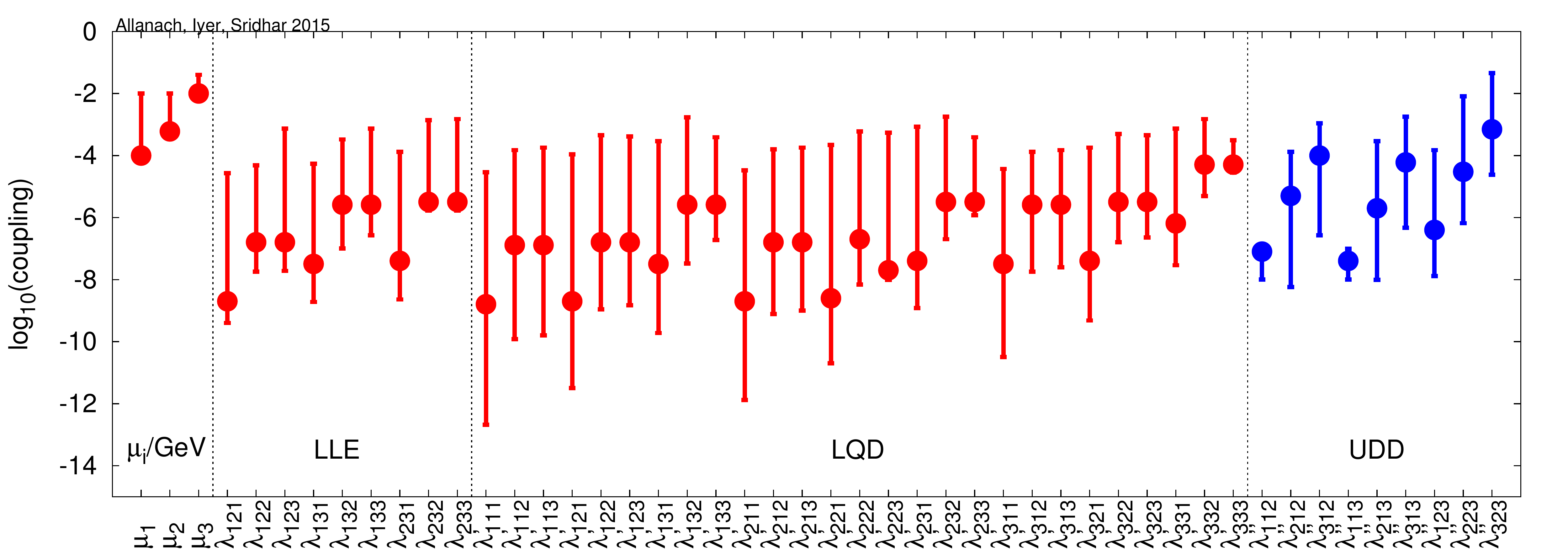}
\caption{\label{RPVparameters} Pattern of $\slashed{R}_p$ couplings. The
  vertical bars give the range of couplings that result from good fits to
  fermion masses and mixings for $\tan \beta=5$, and that respect experimental
  bounds on 
  $\slashed{R}_p$ couplings (if either only the baryon number violating or
  lepton-number violating couplings are allowed). The points correspond to a
  pattern from one particular fit (see sections~\ref{lnv},\ref{bnv} for details).}
\end{figure}
Fig.~\ref{RPVparameters} gives the ranges of the
$\slashed{R}_p$ couplings predicted from the good-fit scanned points, and
constitutes the main result of the present
paper. Dimensionless $\slashed{R}_p$ couplings that are larger than around
$10^{-6}$ result in 
prompt decays of the lightest supersymmetric particle at colliders, whereas if
all couplings are smaller than $10^{-6}$, displaced couplings result. We note
that the smallest couplings are always predicted to be larger than this lower
limit and so $\slashed{R}_p$ decays are prompt. 

We note from Fig.~\ref{RPVparameters} that the $\lambda'_{ijk}$
couplings have a possibility to be smaller than the $\lambda_{ijk}$ and the
$\lambda''_{ijk}$ couplings. This 
can be attributed to the fact that, in the latter case, the couplings are
separately determined by the fits to the lepton and quark sector. As a result
the individual $c_i$ parameters in each sector are interlinked so as to
reproduce the correct hierarchy in the mass matrix. For instance, given a
choice $c_{L_1}$, there is less freedom in the choice of $c_{L_{2,3}}$. 
The $\lambda'_{ijk}$ couplings on the other hand, depend on $c_{L_i},c_{Q_j}$
and $c_{D_k}$. Thus, for
a given choice of $c_{Q_j}$, $c_{D_k}$, which are related from the quark mass
fits, there 
is freedom in the choice of $c_{L_i}$ which are determined from the fits to
leptonic sector and are decoupled from the quark sector at tree level. 
We now proceed to discussing the phenomenological implications of the presence
of these $\slashed{R}_p$ parameters. 
 		
 		\subsection{Nucleon decay}
\begin{figure}
% \picture(150,100)(0,-15)
% \GOval(10,45)(15,70)(90){0.9}
% \GOval(140,0)(15,30)(90){0.9}
% \Text(5,90)[c]{$u$}
% \Text(5,10)[c]{$d$}
% \Text(5,-10)[c]{$u$}
% \Text(10,50)[c]{$p$}
% \Text(140,0)[c]{$\pi^0$}
% \Text(75,58)[c]{$\tilde d_j$}
% \Text(145,90)[c]{$e^+$}
% \Text(145,10)[c]{$\bar u$}
% \Text(145,-10)[c]{$u$}
% \ArrowLine(10,10)(50,50)
% \ArrowLine(10,90)(50,50)
% \DashLine(50,50)(100,50){5}
% \ArrowLine(140,90)(100,50)
% \ArrowLine(140,10)(100,50)
% \ArrowLine(10,-10)(140,-10)
\includegraphics[width=0.4 \textwidth]{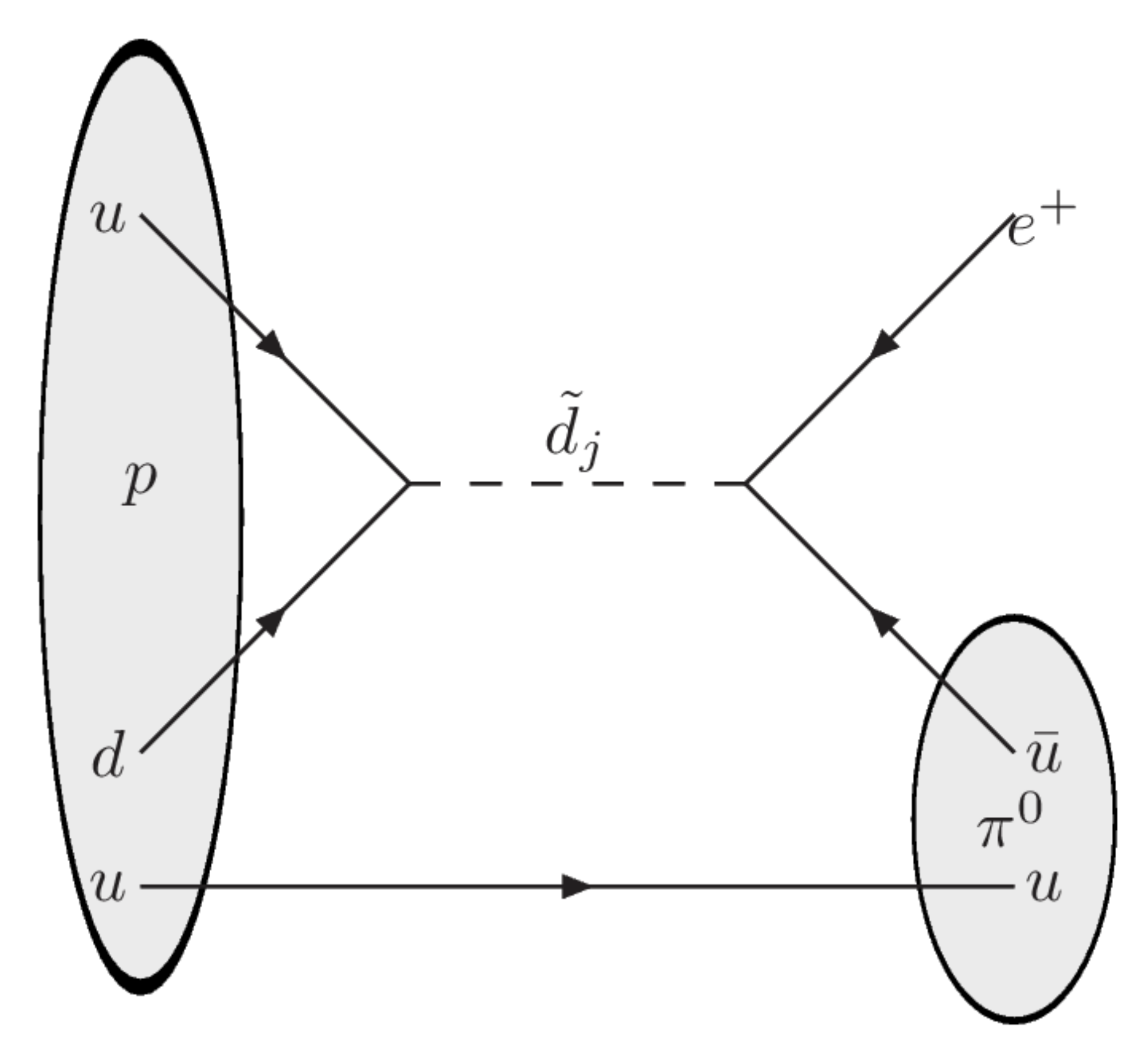}
\caption{Possible $\slashed{R}_p$ violating process ($p \rightarrow \pi^0
  e^+$) yielding a non-zero decay
  rate for non-zero $\lambda'_{11j}\lambda''_{11j}$. \label{fig:pdk}}
\end{figure}
The presence of both lepton- and baryon-number violating terms
in the lagrangian can give rise to small proton-decay
lifetimes for baryon and lepton number violating couplings
being simultaneously non-zero. For instance, a combination of
$\lambda''_{ijk},\lambda'_{ijk}$ can give rise to the
contribution to proton decay shown in Fig.~\ref{fig:pdk}.
This leads to particularly stringent constraints on the sizes of the
couplings. Some of the strongest constraints come from searches for the
following decays~\cite{Eidelman:2004wy}:
 		\begin{eqnarray}
 			|\lambda'_{l1k}\lambda^{''*}_{11k}|&\leq&
                        2\times10^{-25}\left(\frac{\tilde m}{\textrm{1 TeV}}\right)^2~(l=1,2) ~p\rightarrow[\pi^0l^+]\nonumber\\
 			|\lambda'_{31k}\lambda^{''*}_{11k}|&\leq&
                        7\times10^{-25}\left(\frac{\tilde m}{\textrm{1 TeV}}\right)^2 ~n\rightarrow[\pi^0\bar\nu]\nonumber\\
 			|\lambda'_{i2k}\lambda^{''*}_{11k}|&\leq&
                        3\times10^{-25}\left(\frac{\tilde m}{\textrm{1 TeV}}\right)^2 ~p\rightarrow[K^+\bar\nu].
 			\label{rpvbounds}
 		\end{eqnarray}
 		There exist similar bounds on the product of lepton and baryon
                number violating 
                couplings from other decay modes of the proton and
                neutron~\cite{Rajagopal:1990yx,Chang:1996sw,Bhattacharyya:1998dt,Bhattacharyya:1998bx}.    

We note that, even with a relatively heavy sparticle spectrum at around 1 TeV, 
the product of the minimum values of the couplings in Fig.~\ref{RPVparameters}
will violate the bounds on nucleon decay. The violation of the bounds is
more severe for the couplings involving third generation fermions. 
 		If one insists on simultaneous lepton- and baryon-number
                violation, then a choice of
                $\hat\lambda_{ijk},\hat\lambda'_{ijk},\hat\lambda''_{ijk}\sim
                \mathcal{O}(1)$ is 
                no longer viable. Assuming one has a common scaling factor 
                for each $\slashed{R}_p$ coupling, putting it equal to at most
 		$\delta=2\times 10^{-4}$ is necessary to completely evade
                all the bounds from nucleon decay. Thus the nucleon  decay
                problem is vastly ameliorated, but not solved, by warping. 
 		As a result, we do not pursue this further and instead
                focus on cases where either baryon-number {\em or}\/
                lepton-number is 
                violated, predicting a stable proton and none of
                the dangerous lepton and baryon number violating nucleon decay
                channels

 		\subsection{Lepton  Number Violation Only\label{lnv}}
 		Since baryon-number is conserved, the proton does not decay in
                this case. However in this scenario there are
                additional contributions to the neutrino mass: tree level
                contributions originating from
                $\mu_i$~\cite{Joshipura:1994ib}, and loop-induced
                contributions from
                $\lambda_{ijk},\lambda'_{ijk}$~\cite{Hall:1983id,Babu:1990fr}. 
                While 
                it may be possible to 
                generate $\mathcal{O}(0.1)$ eV neutrino masses  with
                these couplings, it is very difficult to satisfy the solar and
                atmospheric neutrino data, which require neutrino mass
                splittings one or two orders of magnitude smaller than this. 
 		As a result, we focus on the parameter space where these
                contributions are suppressed in comparison to 
 		Dirac neutrino mass terms generated by
                Eq.~\ref{fermionmass}. Non-supersymmetric Randall-Sundrum
                scenarios in which lepton-number violating effects could be
                hidden have been
                considered in Refs.~\cite{Gherghetta:2003hf,Iyer:2013hca}. 
 		
 		While it may be simple to suppress the masses of the electron
                and muon neutrinos,
                $\lambda'_{333}$ (being relatively large compared to the other
                $\slashed{R}_p$ couplings) and give rise to too
                heavy a tau neutrino as compared to data. We focus on the
                following region of parameter space, which leads to
                $\slashed{R}_p$ contributions to neutrino
                masses that are not larger than the observed mass splittings:
 		\begin{eqnarray}
 		\lambda'_{133}<10^{-6},\qquad \lambda'_{233}<5\times
                10^{-6},\qquad\lambda'_{333}<5\times 10^{-6}, \qquad
                \mu_3\lesssim 0.01~\text{GeV}. 
 		\label{condition1}
 		\end{eqnarray}
 		The condition Eq.~\ref{condition1} requires
                the lepton doublets to be  
 		far away from the IR brane. However in order to
                fit the required charged lepton masses, the SM singlets must
                then 
                be localized relatively close to the IR brane.
                Picking one particular `good-fit' point, we have:
 		\begin{eqnarray}
 		c_{L_1}= 2.32,~~~c_{L_2}= 2.27,~~~c_{L_3}= 1.61,~~~c_{E_1}= 1.74,~~~c_{E_2}= 0.5,~~~c_{E_3}= 0.5.
 		\label{leptoncvalues}
 		\end{eqnarray}
 		The corresponding values of the lepton-number violating
                couplings in this case are represented by the points in
                Fig.~\ref{RPVparameters}. 
 		One may push the lepton doublets to be further away from the
                IR brane by choosing $c_{E_i}<0.5$. However, this choice is not
                ideal as this may induce large off-diagonal elements in the
                slepton mass matrix, potentially leading to
                large (and excluded) flavour violation.  
 		Along with  Eq.~\ref{leptoncvalues}, one example of a good-fit
                point includes the following choices:
 		\begin{eqnarray}
 		c_{Q_1}= 0.68,~~c_{Q_2}= 1.04,~~c_{Q_3}= 0.77,~~c_{D_1}=
                2.89,~~c_{D_2}= 2.07,~~c_{D_3}= 1.48,\nonumber\\c_{U_1}=
                3.5,~~c_{U_2}= 1.98,~~c_{U_3}=0.47,~~M_1= M_2=
                2.5\text{~TeV},~~M_3= 1.2~\text{TeV}. 
 		\label{quarkcvalues}
 		\end{eqnarray}
        %         The $\slashed{R}_p$ parameters which correspond to the choice of $c_i$
        %         parameters in Eqs.~\ref{leptoncvalues},\ref{quarkcvalues} are
        % represented by points in the red lines of Fig.~\ref{RPVparameters}
        % when the $\mathcal{O}(1)$ parameters  in Eq.~\ref{condition1} and
        %~\ref{condition2} are all set to be to be 1. 
 	 With this choice, the masses of the neutrinos at the low scale can be
         determined using {\tt{SOFTSUSY}}
         \cite{Allanach:2001kg,Allanach:2011de} and are predicted to be:
 		\begin{eqnarray}
 		m_{\nu_1}=1.6\times 10^{-6}~\text{eV};~~~m_{\nu_2}=7.4\times
                10^{-6}~\text{eV};~~~m_{\nu_3}=0.8~\text{eV}. 
 		\end{eqnarray}
 		These masses do respect the direct constraints upon neutrino
                masses, however they do not respect oscillation data, which
                require $\Delta m^2_\text{atm} \sim 2 \times 10^{-3}$ eV$^2$ and
                $\Delta m^2_\text{sol} \sim 7.5 \times 10^{-5}$ eV$^2$ to be
                the values of 
                differences in the neutrino masses
                squared~\cite{Maltoni:2004ei}.   

                In order to suppress the $\slashed{R}_p$ contribution to the
                neutrino masses, we make the following choices:
                \begin{eqnarray}
                  \hat\lambda'_{133}=0.1;~~~\hat\lambda'_{233}=0.1;~~~\hat\lambda'_{333}=0.2;~~~\hat{\mu}_3=0.1
                  \label{condition3}
                \end{eqnarray}
                With this choice, the $\slashed{R}_p$ violating contributions
                to the neutrino masses are  then 
                \begin{eqnarray}
                  m_{\nu_1}=1.0\times 10^{-8}~\text{eV},~~~m_{\nu_2}=4.0\times 10^{-6}~\text{eV},~~~m_{\nu_3}=0.008~\text{eV},
 			\end{eqnarray}
smaller than the values required to satisfy oscillation data.
                        (We could also have suppressed the $\slashed{R}_p$
                        contribution by further raising the
                         gaugino masses $M_{1}, M_2$ from the values in
                         Eq.~\ref{quarkcvalues} at the expense of making
                         the supersymmetric spectrum heavy, thus worsening the
                         supersymmetric solution to the hierarchy
                         problem).

In addition to the operators in
                        Eq.~\ref{RPVlagrangian}, operators of the form
                        $(L_iH_u)(L_jH_u)$ can also contribute to neutrino
                        masses. 
 			This operator violates lepton number by 2 units.
                        The superpotential term is given  by
                         \begin{equation}
 			\mathcal{W}_{\Delta L=2}=\frac{\kappa_{ij}}{M_{Pl}}(L_iH_u)
                        \cdot (L_jH_u),
 			\end{equation}
 			where $\kappa_{ij}$ are 5D Yukawa couplings with mass
                        dimension $M^{-1}$. The neutrino mass matrix entry
                        generated from 
                        this operator is given by
\begin{equation}
(m_\nu)_{ij}=\hat\kappa_{ij}\frac{v_u^2}{2M_{Pl}}e^{kR\pi}f(c_{L_i})f(c_{L_j})
\end{equation}
where $\hat \kappa_{ij}=k \kappa_{ij}$ is a dimensionless $\mathcal{O}$(1)
parameter and the 
function $f$ is defined in Eq.~\ref{wavefunction}. 
 For $c_{L_i}=c_{L_j}>0.5$, this expression can be simplified to
 \begin{equation}
 (m_\nu)_{ij}\sim \frac{v_u^2}{2M_{Pl}}e^{(2-2c_{L_i})kR\pi},
 \end{equation}
which for $c_{L_i}=c_{L_j}=1.6$ comes out to be around $10^{-5}$ eV, 
much smaller than the Dirac mass contribution from Eq.~\ref{fermionmass}.
The $(L_iH_u)(L_jH_u)$ contribution is generally negligible in our model. 	

                \begin{figure}[here]
                  \begin{tabular}{cc}
                    \includegraphics[width=8cm]{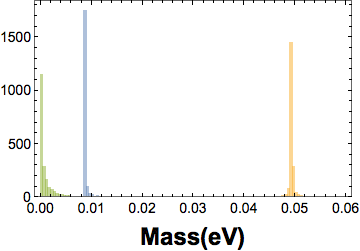}& 				\includegraphics[width=8cm]{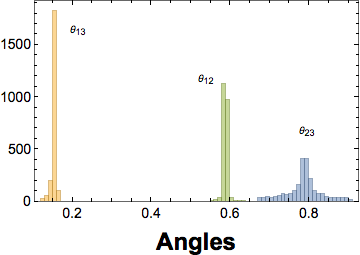}\\
                  \end{tabular}
                  \caption{The left hand plot shows the Dirac neutrino mass
                    eigenvalues (eV) predicted by wave function overlap in RS
                    models (a normal hierarchy is assumed). The right hand
                    plot displays the predicted mixing angles. The vertical
                    axis is effectively arbitrary, showing the frequency of
                    the prediction in a large number of scanned models.} 
                  \label{neutrinomassRS}
                \end{figure}
We are now free, after the addition of right-handed neutrino superfields, 
to arrange for dominant Dirac
                       contributions to the neutrino masses. 
	{The oscillation parameters are determined from the fits to the leptonic data as outlined in Section~\ref{section2}.
		The $c_{N_i}$ parameters (for the right handed neutrinos) which pass the filtering criteria give rise to specific forms of neutrino mass textures leading	to a determination of the neutrino parameters. The mixing angles and the mass eigenvalues can be determined by using the $c_{N_i}$
		in Eq.(\ref{fermionmass}).  Corresponding to the set in Eq.(\ref{leptoncvalues}), the  set of $c_{N_i}$ parameters is:
		\begin{equation}
		c_{N_1}=6.26~~~~~~c_{N_2}=5.99~~~~~~c_{N_3}=8.72
		\end{equation}	
%                Predictions in line with oscillation data are easy to achieve
%                in a RS model that generates the masses purely from wave
%                function overlap in the Dirac masses, as shown in 
%                Fig.~\ref{neutrinomassRS}. 
                Fig.~\ref{neutrinomassRS} shows the results of the fit to the neutrino oscillation data obtained from the model parameters.
	         	The left-handed plot shows the
                predicted distribution of 
                neutrino 
                mass eigenvalues using different sets of $c_{N_i}$ parameters satisfying both $\delta m^2_{12}\sim
                7.5\times 10^{-5}~\text{eV}^2$ and $\delta m^2_{32}\sim
                2.32\times 
                10^{-3}~\text{eV}^2$.  In addition, the corresponding PMNS mixing angle
                predictions  are
                shown 
                in the right hand plot of Fig.~\ref{neutrinomassRS}, and they
                are close to the values inferred from
                experiment~\cite{Maltoni:2004ei}. Thus we see that, predictions in line with oscillation data are easy to achieve
                in a RS model that generates the masses purely from wave
                function overlap in the Dirac masses.

 			\begin{table}[htbp]
 				\renewcommand{\arraystretch}{1.5}
 				\begin{center}
 					\begin{tabular}{|c|c|c|c|c|c|c|c|}
 						\hline
 						Parameter&Mass/TeV&Parameter&Mass/TeV&Parameter&Mass/TeV&Parameter&Mass/TeV\\  
 						\hline 
 						
 						$\tilde t_1$ &   1.8&$\tilde b_1$  &   2.2& $\tilde\tau_1$ & 1.1&$\tilde\nu_\tau$ &  1.6\\
 						$\tilde t_2$ &   2.3&$\tilde b_2$  &   2.3&$\tilde\tau_2$ &   1.6&$\tilde\nu_\mu$ &  1.6\\
 						$\tilde c_1$ &   2.2&$\tilde s_1$ &   2.2&$\tilde\mu_R$  &   1.2&$\tilde\nu_e$ &   1.6\\
 						$\tilde c_2$ &   2.7& $\tilde s_2$ &   2.7&$\tilde\mu_L$  & 1.6&$\tilde g$&2.6\\
 						$\tilde u_1$ &   2.2& $\tilde d_1$  &   2.2&$\tilde e_R$   &   1.1&$\chi^\pm_1$&2.0\\
 						$\tilde u_2$ &   2.7&$\tilde d_2$  &   2.7& $\tilde e_L$   &   1.6&$\chi^\pm_2$&2.3\\
 						$m_{A^0}$&3.1&$m_H^{\pm}$&3.1&$m_h$&0.121&$m_H$&3.1\\
 						$\chi^0_1$&1.1&$\chi^0_2$&2.0&$\chi^0_3$&2.3&$\chi^0_4$&2.4\\
 						\hline 
 					\end{tabular}
 				\end{center}
 				\caption{Example supersymmetric spectrum for
the lepton number violating case and $\tan \beta=5$. }
 				\label{spectrum}
 			\end{table}
The supersymmetric
                         spectrum corresponding to the choice of GUT scale
                         parameters\footnote{It was shown in
                           \cite{Cohen:2006qc} that the running of soft masses
                           may depend on physics in the hidden sector which
                           breaks SUSY\@. These effects are likely to be
                           relevant only 
                           for third generation squarks and are not included
                           here because they add additional model dependence }
                         obtained from
                         Eqs.~\ref{leptoncvalues},\ref{quarkcvalues} is given
                         in Table~\ref{spectrum}.  
The
lightest CP-even Higgs mass $m_h$ is predicted a little on the low side, but the
discrepancy with the experimental measurement of 0.125 TeV can be explained by
missing higher order corrections in its prediction. 
The spectrum is heavy enough to have not yet been ruled out by LHC
constraints, but light enough to expect a discovery in future runs. 
Indeed, the model predicts that there will be many multi-lepton rich signals
from all of the lepton-number violating couplings that are switched on:
strongly interacting sparticles are likely to be detected first. These then
undergo cascade decay via $R_p$ conserving processes until the lightest
supersymmetric particle - in this case the $\tilde e_R$ - is reached: this is
because the $R_p$ preserving dimensionless couplings such as gauge couplings
and third family Yukawa couplings are larger than the $\slashed{R}_p$ ones
that are shown in Fig.~\ref{RPVparameters}.
This particle then decays via $\slashed{R}_p$: the predominant decay
in this case is via $\lambda'_{133}$ 
into a bottom quark and an anti-top. 
Thus, SUSY events are $b$-rich (predicting 4 $b$ quarks) and may produce
leptons from top decays, or be susceptible to top taggers. 
There is a non-zero branching ratio for $\tilde e_R \rightarrow \mu 
\nu_\tau$ via a similar sized coupling $\lambda_{132}$, and the additional
muons may also aid detection strategies in multi-lepton channels. 

We note that recent $\slashed{R}_p$ explanations of an apparent excesses 
in LHC data~\cite{Allanach:2014lca,Allanach:2014nna,Allanach:2015blv,Ding:2015rxx,Allanach:2015ixl} are not
naturally accommodated in this 
set-up. They all are based on resonant slepton production and require a large
order 0.1 coupling $\lambda'_{i11}$, which 
is not possible in our set-up. One would need additional flavour symmetry in
order to fix some $\slashed{R}_p$ couplings to zero and reduce the effect of
various bounds on products of them in order to be able to accommodate such a
coupling.

 	\subsection{Baryon-number violation only ~\label{bnv}}
 	We now consider a scenario where only baryon-number violating terms are included in the lagrangian.
 	Since lepton number is perturbatively conserved in this case, the
        superpotential terms proportional to 
        $\lambda_{ijk}$, $\lambda'_{ijk}$ or $\mu_i$  are absent. Proton decay
        is forbidden 
        as it requires the presence of both baryon- and lepton-number
        violating terms in the Lagrangian. The neutrinos in this case must be
        purely Dirac type and their masses are determined using
        Eq.~\ref{fermionmass}, just as for the other charged fermions. 
 	The results of the fit to the neutrino oscillation data is given in
        Fig.~\ref{neutrinomassRS}. 

We illustrate the spectrum for the following parameter choice: 	
 \begin{eqnarray}
 	c_{Q_1}&=& 2.2,~~c_{Q_2}= 1.7,~~c_{Q_3}= 0.7,~~c_{D_1}= 1.8,~~c_{D_2}= 1.2,~~c_{D_3}= 1.4,\nonumber\\c_{U_1}&=& 2.3,~~c_{U_2}= 1.3,~~c_{U_3}= 0.3,~~c_{L_1}= 2.2,~~c_{L_2}= 1.8,~~c_{L_3}= 1.4,\\\nonumber c_{E_1}&=& 1.7,~~c_{E_2}= 0.9,~~c_{E_3}= 0.5,~~M_1= 1.0\text{~TeV} ,~M_2= 1.0\text{~TeV},~M_3= 1.4\text{~TeV}.
 	\label{cvaluesbnv}
 \end{eqnarray}
 The choice of the corresponding $\slashed{R}_p$ couplings is represented by the
 blue points in Fig.~\ref{RPVparameters}. 
 Note that the lepton doublet fields need not be so strongly localized towards
 the UV as in the lepton number violating case because there are no bilinear
 couplings which can contribute to the neutrino masses. This is reflected in
 the values of $c_{L_i}$. 
 Table~\ref{spectrumbnv} gives the low energy spectrum corresponding to the
 choice of GUT scale parameters by Eq.~\ref{cvaluesbnv}. 
 	\begin{table}
 		\renewcommand{\arraystretch}{1.5}
 		\begin{center}
 			\begin{tabular}{|c|c|c|c|c|c|c|c|}
 				\hline
 				Parameter&Mass/TeV&Parameter&Mass/TeV&Parameter&Mass/TeV&Parameter&Mass/TeV\\  
 				\hline 
 				
 				$\tilde t_1$ &   2.3&$\tilde b_1$  &   2.3& $\tilde\tau_1$ & 0.3&$\tilde\nu_\tau$ &  0.3\\
 				$\tilde t_2$ &   2.7&$\tilde b_2$  &   2.8&$\tilde\tau_2$ &   1.0&$\tilde\nu_\mu$ &  0.3\\
 				$\tilde c_1$ &   2.8&$\tilde s_1$ &   2.8&$\tilde\mu_1$  &   0.3&$\tilde\nu_e$ &   0.3\\
 				$\tilde c_2$ &   2.7& $\tilde s_2$ &
                                2.8&$\tilde\mu_2$  &   0.9&$\tilde g$&3.2\\
 				$\tilde u_1$ &   2.8& $\tilde d_1$  &   2.8&$\tilde e_1$   &   0.3&$\chi^\pm_1$&0.8\\
 				$\tilde u_2$ &   2.6&$\tilde d_2$  &   2.8& $\tilde e_2$   &   0.9&$\chi^\pm_2$&3.1\\
 				$m_{A^0}$&3.3&$m_H^{\pm}$&3.3&$m_h$&0.121&$m_H$&3.3\\
 				$\chi^0_1$&0.1&$\chi^0_2$&1.0&$\chi^0_3$&2.0&$\chi^0_4$&2.1\\
 				\hline 
 			\end{tabular}
 		\end{center}
 		\caption{Example supersymmetric spectrum for 
                         $\tan \beta=5$ in the baryon number violating case.} 
 		\label{spectrumbnv}
 	\end{table}
 We find the spectrum has a nice feature wherein the coloured sparticles are
 grouped together in a small mass window. The sleptons in this case have a
 tendency to be lighter that the lepton number violating case as the there are
 no constraints coming from  upper bounds on the neutrino masses. The
 light smuon and neutralino gives a non-negligible contribution to the
 anomalous magnetic moment of the muon $(g-2)_\mu$, which may explain the
 apparent 
 3.6$\sigma$ discrepancy between measurements and SM predictions: $\delta
 (g-2)_\mu/2=(29 \pm 8) \times 10^{-10}$~\cite{Bennett:2006fi}. SUSY loops
 with smuons and 
 neutralinos running in the loop yield $\delta (g-2)_\mu/2\approx 13 \times
 10^{-10}  \left(100\text{~GeV}/\text{max}(m_{{\tilde \mu}_L},\
   m_{\chi_1^0})\right)^2 \tan \beta$~\cite{Czarnecki:2001pv} \footnote{For recent work on explaining $g-2$ in $\slashed{R}_P$ SUSY see \cite{Chakraborty:2015bsk}}. Thus, it
 appears that by 
 increasing $\tan \beta$ (which may go as high as 50) one may fit $(g-2)_\mu/2$.
Again, the spectrum presented is allowed by previous collider constraints, but
should be covered in coming LHC runs. Again, production of the strongly
interacting particles will proceed via $R-$parity conserving decays, and
usually end in the lightest neutralino $\chi_1^0$. This will then decay via
$\lambda''_{323}$ into a top, a strange and a bottom so we again expect
bottom-rich events (at least four), but now there is no obvious source of
missing energy unless leptons come from the top decay with an associated
neutrino. The `golden' decay chain $\tilde q \rightarrow \chi_{2}^0 q
\rightarrow \tilde e e q \rightarrow \chi_1^0 e^+ e^- q$ is also open, which
may lead to interesting invariant mass edges between the leptons (golden decays
with $e$ replaced by $\mu$ in the preceding decay should also be present). 

\section{Conclusions}

In a general supersymmetric extension of the SM, lepton and baryon number are
not necessarily perturbatively conserved, unless a symmetry such as $R-$parity
is invoked.
As a result, the most general supersymmetric lagrangian includes terms
which violate both these symmetries. This however increases the number of free
parameters in the form of undetermined values of the $\slashed{R}_p$ couplings. 

In this work we propose a scenario by embedding the MSSM in a higher
dimensional warped framework. Following the aesthetic that all dimensionless
parameters should be of order 1 in a fundamental theory
the warped dimensional set-up provides the flavour
structure, while supersymmetry resolves the technical hierarchy problem. 
All of the  supersymmetric parameters at the GUT
scale including the $\slashed{R}_p$ couplings
are determined by the same set of parameters which fix the fermion masses and
mixings 
at this scale. This lends a certain level (order of magnitude-wise) of
predictability to the framework, and we present, in Fig.~\ref{RPVparameters},
predictions of ranges of $R-$parity violating parameters. The couplings
involving the third family tend to be the largest because of the warping
structure. 
The predictions
typically range over several orders of magnitude but are dependent on the
flavour indices of the coupling.
  	
 In the most general scenario which includes both baryon and lepton number
 violating terms, the nucleon decays too quickly for $\sim {\mathcal O}(1)$
 dimensionless $\slashed{R}_p$ parameters, although if instead
 they are all set to be 
 $\mathcal{O}(10^{-4})$, the lifetime may be long enough to evade current
 experimental bounds (for
 superpartner masses of around 2 TeV). Following our initial idea of the
 aesthetic, it appears though that one needs to
 forbid either the lepton-number or baryon-number violating terms, in which
 case plenty of parameter space exists where current experimental bounds on
 the couplings are respected.

For the case where only lepton
 number is violated we find points in parameter space where the neutrino
 masses are predominantly Dirac-like nature, even in the presence of various
 lepton number violating operators contributing to the neutrino masses. 
The neutrino masses and mixings are fit to oscillation data just as the
charged fermions are fit.
In the baryon number violating case, the sleptons have a 
 tendency to be lighter making it more appealing from the collider searches
 point of view: leptons may appear more often in supersymmetric decay chains,
 providing clean objects with low backgrounds to search for. In addition, the
 lighter smuons mean that a supersymmetric explanation for the discrepant
 anomalous magnetic moment of the muon is viable. 
In either the lepton-number or baryon-number violating cases, LHC signals
consist of prompt hard jets, and $b-$rich events (at least four per event are
predicted) containing tops. 
In the lepton-number violating case there may also
be a modest amount of missing transverse momentum coming from neutrino
production. 
We illustrate points in parameter space where current collider
limits are respected but where the LHC should be able to discover sparticles
in future runs, which we eagerly await. 

 		\section{Acknowledgements}
 		This work was partially supported by STFC grant  ST/L000385/1.
 		BCA was adjunct faculty in DTP TIFR during December 2014 and would like to thank the department for the hospitality extended during the early stages of this work and the Cambridge SUSY Working Group for helpful suggestions.
 		BCA, AI and KS would like to thank Sudhir Vempati for the motivation  and useful discussions. AI and KS also thank the organisers at WHEPP 2013 for the hospitality where the idea was conceived. AI and KS would also like to thank the CERN theory division for hospitality where part of the work was completed.
 		AI would like to thank Tuhin Roy for discussions on contribution to renormalization of soft masses due to hidden sector effects.\\
 		
 {\em AI would like to dedicate this paper to the memory of close friend and fellow physicist \\  Dr. Senti Imsong, who passed away recently.}	
 		
 		\bibliographystyle{apsrev}
 		
 		\bibliography{rpv.bib}

 	\end{document}